\documentclass[12pt,aps,prb,superscriptaddress]{revtex4-1}
\usepackage[dvips]{graphicx}
\usepackage{amssymb,amsfonts,amsmath}
\usepackage[sort&compress]{natbib}
\usepackage{color}
\usepackage[outdir=./]{epstopdf}

\begin{document}

\title{Programmable Active Janus Droplets Driven by Water/Alcohol
  Phase Separation}

\author{Menglin Li} \affiliation{Experimental Physics, Saarland
  University, 66123 Saarbr\"ucken, Germany}

\author{Martin Brinkmann} \affiliation{Experimental Physics, Saarland
  University, 66123 Saarbr\"ucken, Germany} \affiliation{Max Planck
  Institute for Dynamics and Self-Organization, 37077 G\"ottingen,
  Germany}

\author{Ignacio Pagonabarraga} \affiliation{Department of Condensed
  Matter Physics, University of Barcelona, Carrer de Marti i Franques
  1, Barcelona, Spain}

\author{Ralf Seemann} \affiliation{Experimental Physics, Saarland
  University, 66123 Saarbr\"ucken, Germany} \affiliation{Max Planck
  Institute for Dynamics and Self-Organization, 37077 G\"ottingen,
  Germany}

\author{Jean-Baptiste Fleury}
\email{jean-baptiste.fleury@physik.uni-saarland.de}
\affiliation{Experimental Physics, Saarland University, 66123
  Saarbr\"ucken, Germany}

\date{\today}

\begin{abstract}
  We report the existence of self--propelled Janus droplets driven by
  phase separation, which are able to deliver cargo in a programmable
  manner.  The self--propelling droplets are initially formed by a
  water/ethanol mixture in a squalane/monoolein solution, and evolve
  in up to three stages depending on ethanol concentration. In the
  first stage, the droplet propulsion is generated by Marangoni flow
  originating from the solubilization of ethanol in the oily
  phase. During this process the droplets absorb surfactant molecules;
  in combination with the continuous loss of ethanol this leads to a
  phase separation of the water/ethanol/monoolein mixture and the
  formation of Janus droplets, i.e.~a water--rich droplet connected to
  an ethanol--rich droplet that is able to deliver cargo.  We
  characterize the different evolution stages of self-propulsion by
  the flow field around the droplet that evolves from a weak pusher,
  over a neutral swimmer, to a dimer of neutral swimmers. Finally, we
  utilize this active system to deliver DNA as a cargo. Tuning the
  delay time before phase separation, by varying the chemical
  composition of the droplets, several different cargo delivery
  processes can be programmed.
\end{abstract}


\maketitle


\section{Introduction}
\label{sec:introduction}

 In recent years, significant efforts were dedicated to realize
 artificial micro-- and nano--swimmers that locomote at low Reynolds
 numbers, and that are further able to perform tasks, like cargo
 delivery~\cite{Sanchez2015, Baylis2015, Wang2013}. Swimmers driven by
 periodic conformational changes have to break the time-reversal
 symmetry to achieve a net propulsion at small Reynolds numbers
 \cite{Purcell1977, Lauga2011}. However, swimmers driven by a
 continuous propulsion mechanism need to break spatial symmetry and
 display a permanent polarity. Despite the theoretical possibility to
 observe a spontaneous symmetry breaking for homogeneous spherical
 particles \cite{Michelin2013}, all solid autophoretic swimmers
 realized in experiments so far exhibit an ingrained
 polarity. Spherical Janus particles, for instance, exploit a
 difference in chemical surface composition of the two opposing
 hemispheres \cite{Jiang2010}. Recent studies have shown that Janus
 nanorods can not only self--propel but also deliver drugs during
 self--propulsion~\cite{Xuan2014}, repair cracks in electrical
 circuits~\cite{Li2015a}, or even halt blood
 hemorrhages~\cite{Baylis2015}.  Beside their desired properties, is
 the nature of these colloidal (Janus) particles and the way to
 trigger their self--propulsion often incompatible with biological
 applications~\cite{Sanchez2015}.

In contrast, emulsion droplets can be formed with bio--compatible
liquids. It was shown that emulsion droplets can self--propel by
Marangoni flow~\cite{Seemann2016} when a surface tension gradient is
either maintained by chemical reactions~\cite{Thutupalli2011},
micellar solubilization~\cite{Peddireddy2012, Izri2014}, or
liquid--liquid phase separations~\cite{Maass2015}. However, the
realization of self--propelled Janus droplets has been proven
difficult~\cite{Bormashenko2011, Jeong2015, Choi2013, Guzowski2012,
  Li2015b} and swimming droplets could so far not perform tasks
comparable to solid self--propelling particles.

In this article, we report time--evolutive self--propelled
ethanol/water droplets in a surrounding squalane/monoolein solution,
which are able to deliver cargo. The evolution of the emulsion
droplets is caused by a continuous release of ethanol from the droplet
into the continuous phase and simultaneous uptake of monoolein from
the continuous phase into the droplet phase. This leads, after a
certain time, to a spontaneous phase separation between water and
ethanol in the presence of the surfactant monoolein, see
Fig.~\ref{fig:exp1}a. Depending on initial ethanol/water mixing ratio,
the droplets evolve in up to three stages and potentially form Janus
droplets with a water--rich leading droplet and an ethanol--rich
trailing droplet. We describe and characterize the conditions of
emergence and evolution of the different stages, and their
corresponding propulsion mechanism. Furthermore, we demonstrate that
the droplets can extract and precipitate DNA during the phase
separation process, which can is delivered via the ethanol--rich
droplet to a target location. Because the emergence of phase
separation is controlled by the initial chemical composition of the
droplets, several different cargo delivery processes can be achieved
in a programmable manner. The propulsion mechanism is universal and
could be realized with several other liquid mixtures or lipidic
surfactant, including a wide range of alcoholic beverages (if having
sufficiently high alcohol content).

\section{Results}
\label{sec:results}

\subsection{Production and Hydrodynamic Properties}
\label{subsec:evolutive_droplets}

 The generation of evolutive self--propelling droplets is achieved
 using a four component system consisting of water, a water--miscible
 solvent, which is less polar than water and partially soluble in the
 surrounding phase, an oily phase, and an oil soluble surfactant. The
 surfactant needs a chemical affinity for the solvent and
 the oil should be immiscible with water. A combination of chemical
 substances satisfying all these conditions is water/ethanol/monoolein
 and squalane (as 4\,\% of ethanol is soluble in squalane). For
 simplicity we will restrict our following description to this
 particular system, but the reported concept is universal and similar
 results could be obtained for seven other chemical compositions
 fulfilling the above mentioned requirements, see~\cite{Suppmatt}.

\subsubsection{Evolution stages of the droplets as function of ethanol 
	concentration}
\label{subsubsec:stages}
\begin{figure}[c]
  \centering \includegraphics[width=16cm]{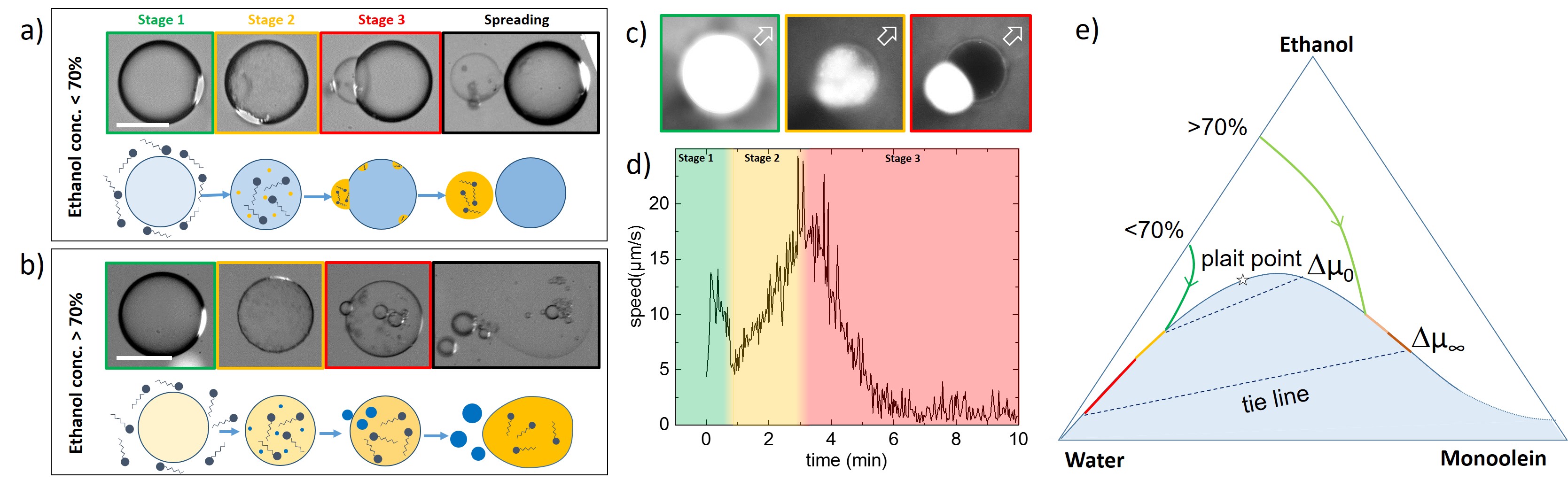}
  \caption{\textbf{Three evolution stages of self-propelling
      water/ethanol emulsion droplets} ethanol solubilization (green),
    nucleation/coalescence (yellow), and Janus (red) for initial
    ethanol concentrations between $30~vol\%$ and $70~vol\%$ (a) and
    for ethanol concentrations $> 70~vol\%$ (b). (Top rows) display
    optical microscopies and (bottom rows) sketches (bottom rows), the
    intensity of the blue and yellow colors indicate increasing
    concentration of water, respectively of ethanol.  c) Optical
    fluorescence micrographs observed for the different droplet stages
    insert. \emph{Stage 1}, (\emph{left}): Droplet appears bright due
    to the fluorescent lipid molecules taken up from the continuous
    phase. For the particular example of $70~vol\%$ ethanol
    concentration additionally dark regions are visible around the
    drop.  \emph{Stage~2}, (\emph{middle}): The fluorescent lipids
    indicate the ethanol--rich phase during nucleation and coarsening.
    \emph{Stage~3}, (\emph{right}): The fluorescent lipids indicate
    that the trailing droplet contains the ethanol--rich phase while
    the leading drop contains the water--rich phase in the Janus
    regime.  d) Droplet velocity as function of time for an initial
    ethanol concentration of $50~vol\%$.  e) Phase diagram of the
    water/ethanol/Monoolein ternary mixture. Two associated pathways
    for two mixtures, one above $70~vol\%$ of ethanol and the other
    below $70~vol\%$, as function of the three evolution stages.}
  \label{fig:exp1}
\end{figure}

Droplets with a diameter of about 200~$\rm\mu$m were generated by
injecting an ethanol/water mixture through a glass capillary (inner
diameter $\approx 20~\mu$m) into a flat observation chamber that is
pre--filled with a continuous squalane/monoolein mixture. The bottom
of the observation chamber is rendered hydrophobic with a
self--assembled monolayer of octadecyltrichlorosilane (OTS)
to prevent adhesion and subsequent spreading of the droplets at the bottom
surface of the chamber. Depending on ethanol concentration, the
generated water/ethanol droplets evolve with time assuming up to three
different stages, as displayed Fig.~\ref{fig:exp1}. We start our
description for a medium ethanol concentration, where all three stages
can be observed (Fig.~\ref{fig:exp1}a), and subsequently extend our
discussion to smaller and respectively to larger initial ethanol
concentrations. In the following discussion the monoolein
concentration in squalane is always fixed to 15~mM, i.e. 7 times above the
critical micellar composition (CMC), if not explicitly stated
otherwise.

For intermediate ethanol concentrations, between $\approx 40~vol\%$
and $\approx 60~vol\%$, the droplets start to self--propel in the oil
phase just after their production. During self--propulsion every
droplet looses a few percent of its volume, and takes up monoolein
molecules from the surrounding phase. The influx of monoolein to the
swimming droplet is monitored by adding fluorescent lipids to the
initial water/ethanol mixture. Similar to monoolein, the fluorescent
lipids display a strong affinity to ethanol and follows the
repartition of monoolein molecules between the fluid phases,
cf. Fig.~\ref{fig:exp1}c.  After $1-3$ minutes in \emph{stage~1}, a
nucleation process occurs inside the self--propelling droplet
(\emph{stage~2}) which lasts for about $2-4$ minutes and leads to the
formation of an ethanol--rich droplet inside a water--rich
droplet. The ethanol--rich droplet is finally pushed out of the
water--rich droplet when touching its interface, generating an
ethanol--rich droplet trailing behind the leading water--rich droplet.
Such a droplet pair formed by the coexistence of two phase--separated
droplets constitutes a long--lived Janus droplet that is
self--propelling for about $5-10$ min (\emph{stage~3}). At the end of
\emph{stage~3}, the two droplets forming the Janus droplet break up,
Fig.~\ref{fig:exp1}a. After separation the ethanol--rich trailing
droplet typically spreads on the hydrophobic substrate and the
water--rich leading droplet stops its motion suddenly, or in rare
cases, continues its self--propulsion by a few droplet radii.

When initially producing water/ethanol droplets with smaller ethanol
concentration of just $\approx 30~vol\%$, the demixing starts almost
immediately after droplet production,
cf. Fig.~\ref{fig:exp1}e. Droplets with this low ethanol concentration
are directly starting to self--propelled in \emph{stage~2}. The
demixing in \emph{stage~2} is very fast ($\approx 1$ min) and the
self--propelling behavior of the formed Janus droplets can effectively
be observed only in \emph{stage~3} for around $5$ min. The qualitative
behavior in these stages is as described previously for intermediate
ethanol concentrations. Reducing the initial ethanol concentration in
the droplets even below $30~vol\%$, phase separation is not always
observed and self--propelled motion is not occurring reliably.

\noindent Droplets produced with an ethanol concentration between
$\approx 70~vol\%$ to $\approx 80~vol\%$ show all three evolution
stages (Fig.~\ref{fig:exp1}b), whereas the duration of \emph{stage~1}
is extended to $10-15$~min before the phase separation occurs in
\emph{stage~2}. The duration of the subsequent demixing in
\emph{stage~2} is around $1-2$ min. In contrast to the intermediate
and low ethanol concentrations, the water--rich phase is now clearly
the minority phase and several water--rich droplets are formed inside
the ethanol--rich droplet. Accordingly, the water--rich droplets are
finally pushed out of the larger ethanol--rich droplet and may
continue to locomote. The ethanol--rich mother droplet finally spreads
on the OTS-coated bottom of the device and stops moving. Generating
droplets with ethanol concentrations well above $80~vol\%$ is not
possible as those droplets directly wet the OTS-coated bottom of the
observation chamber and do not self--propel.

\subsubsection{Characteristics of the evolution stages}
\label{subsec:subsubsection:hydrodynamic_propulsion}

Having introduced the the appearance of the evolution stages as
function of ethanol concentration, we will shed some light on their
characteristics.
%
%
A self--propelling droplet in \emph{stage~1} collects a large amount
of monoolein from the oil phase as can be seen from the fluorescence
activity after a short propulsion time. In case of large ethanol
concentration ($\geq 70~vol\%$), additional dark clouds can be
observed around the droplet, which we suspect to be Marangoni
rolls~\cite{Schwarzenberger2014}, together with a dark trail
(Fig.~\ref{fig:exp1}c).  The droplet is releasing more material with
increasing ethanol concentration, visibly by the reduction of droplet
volume ranging from only a few percent at $30~vol\%$ ethanol
concentration to about $40~vol\%$ for an ethanol concentration of
$80~vol\%$. Obviously, the released material is ethanol which is
soluble by $4~vol\%$ in the surrounding squalane. Because we neither
observed self--propelling nor volume loss in the case of pure water
droplets within the duration of the experiment irrespectively of the
monoolein concentration, we further conclude that the fraction of
water leaving the droplets can be neglected with respect to the
ethanol loss. The absence of self--propelled motion of pure water
droplets agrees with the findings of Thutupalli et
al.~\cite{Thutupalli2011} but is at variance with the observations of
Izri et al.~\cite{Izri2014}.

\begin{figure}[c]
  \centering
  \includegraphics[width=15cm]{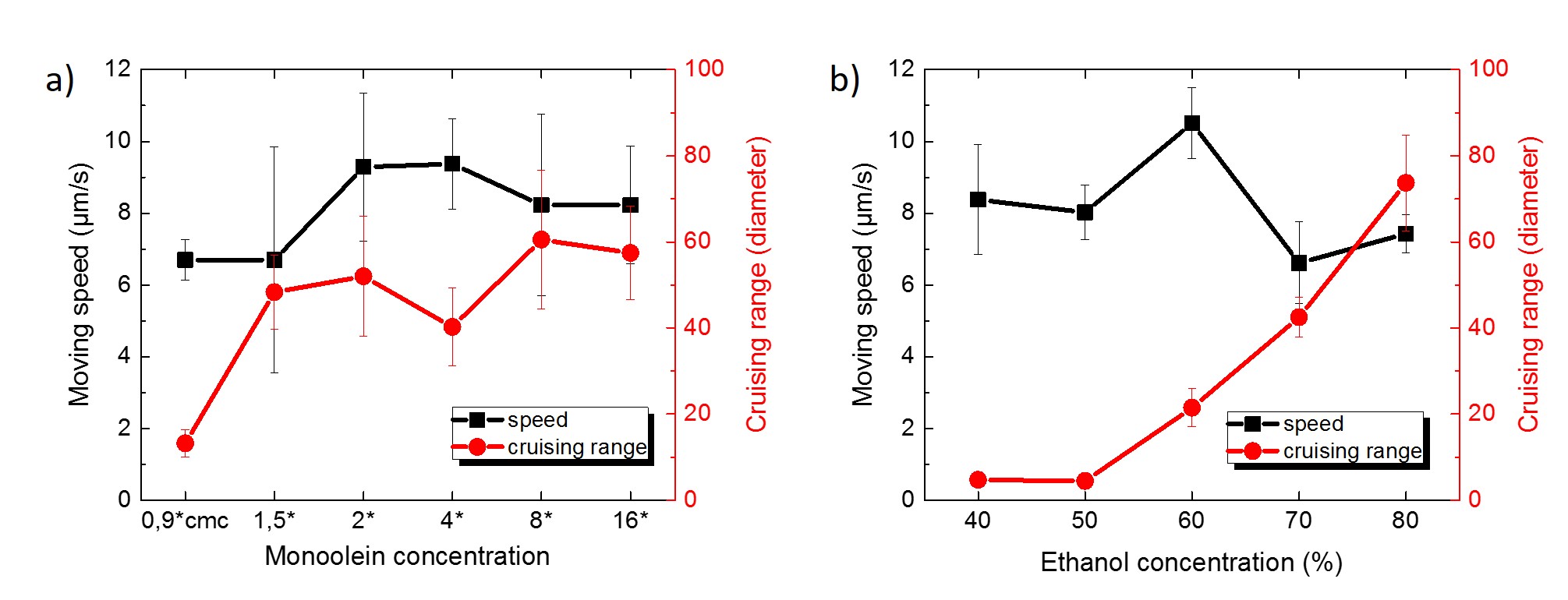}
  \caption{\textbf{Dependence of average droplet velocity and cruising
      range as a function of monoolein (a) and ethanol concentration
      (b) in stage 1.} Cruising range is given in units of the droplet
    diameter. (Note that the total cruising ranges for all three
    stages is typically 200--300 droplet diameter.) The data for
    variable monoolein concentration correspond to an initial ethanol
    concentration of $50~vol\%$; the data for variable ethanol
    concentration correspond to a monoolein concentration of 7 times
    the CMC.}
\label{fig:exp2}
\end{figure}

Based on the above findings we conclude that the solubilization of
ethanol into the squalane solution is driving the droplet motion. This
effect is described for the case of liquid--crystal
droplets~\cite{Peddireddy2012, Maass2015}, where liquid--crystal
molecules are solubilized in a monoolein/squalane solution by filling
surfactant micelles~\cite{Suppmatt}, leading to a variation of the
surface tension at the droplet surface and so to a Marangoni
flow~\cite{Peddireddy2012, Maass2015, Seemann2016}. However our
situation is more complex than in Refs.~\cite{Peddireddy2012,
  Maass2015} as ethanol can be solubilized not only in micelles, but
also at the molecular lever.  We demonstrate this point by observing
the self--propulsion of droplets below the critical micellar
concentration (CMC), cf.~Fig.~\ref{fig:exp2}a , in contrast to all the
other reported examples of self--propelling
droplets~\cite{Peddireddy2012, Thutupalli2011, Izri2014, Maass2015,
  Seemann2016}.

\begin{figure}[c]
  \centering \includegraphics[width=15cm]{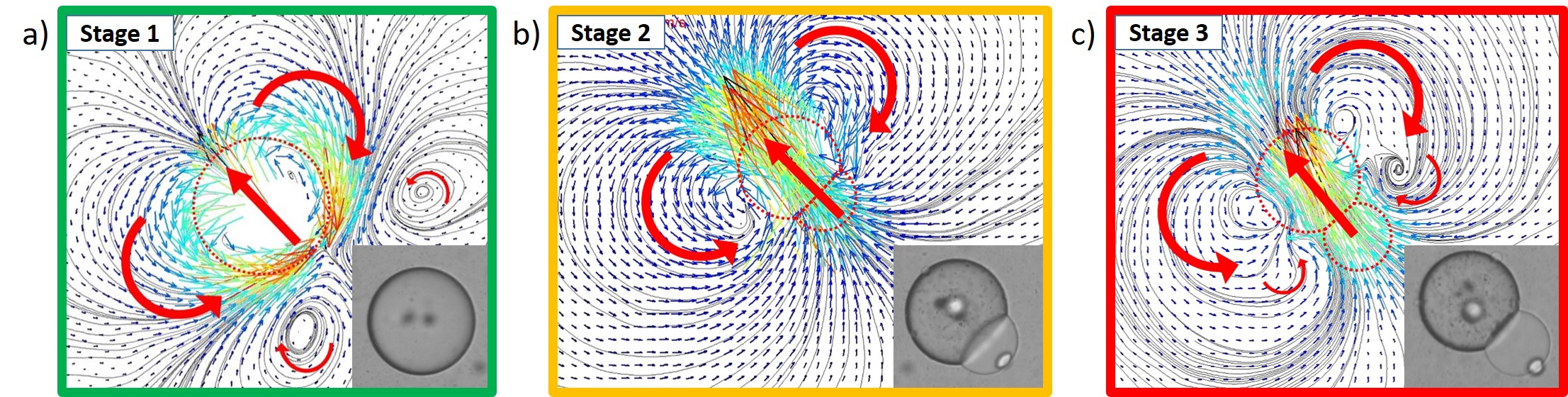}
  \caption{\textbf{Flow fields around the self--propelling droplets in
      the different stages} The flow field are determined by $\mu$PIV
    in the lab frame. The droplets move close to the bottom surface of
    the device, that is at around $2$ times higher than the average
    droplet diameter.  The flow fields were determined parallel to the
    bottom walls and are similar to the rotational symmetric flow
    fields of (a) a weak pusher in \emph{Stage~1}, (b) a neutral
    swimmer in \emph{stage~2}, and (c) a chain of neutral swimmers in
    \emph{stage~3}:. }
  \label{fig:exp3}
\end{figure}

The droplets velocities in \emph{stage~1} are constant in average with
strong fluctuations, similar to other reported systems displaying
self--propelled motion due to solubilization~\cite{Peddireddy2012,
  Maass2015}. The droplet velocity has no clear dependence on the
initial ethanol concentration provided it is in a range that droplets
self--propel ($> 30~vol\%$) and can be stabilized against coalescence
($< 80~vol\%$), see Fig~\ref{fig:exp2}b. But the self--propulsion
running time and with it the cruising range in \emph{stage~1} is
significantly increased with increasing ethanol concentration which is
the fuel for this propulsion process,
cf.~Fig.~\ref{fig:exp2}b. Measurements of the surrounding flow field
by micro-Particle Image Velocimetry ($\mu PIV$), shown in
Fig~\ref{fig:exp3}a, characterize this type of self--propelling
droplets in \emph{stage1} as squirmer in a weak--pusher mode. Note
that this flow field is different from flow fields reported for
comparable systems which reveal a more symmetric flow field, as
expected of neutral squirmers ~\cite{Lighthill1952, Blake1977}

%
Upon self--propulsion, ethanol solubilization and surfactant
absorption are working together until the condition of phase
separation is eventually reached~\cite{Efrat2007, Spicer2001,
  Engstrom1998}, leading to the observed water/ethanol phase
separation in \emph{stage~2}. \emph{Stage~2} persists only a short
time with a continuous increase of the droplet velocity while
nucleation and coarsening processes are ongoing, see
Fig.~\ref{fig:exp1}c. According to the measured flow field around a
droplet in \emph{stage~2}, see Fig.~\ref{fig:exp3}b, the droplets can
be characterized as neutral squirmers. The demixing of water/ethanol
in the presence of monoolein is reported in literature for the
corresponding ternary mixture
(water/ethanol/monoolein)~\cite{Efrat2007, Spicer2001,
  Engstrom1998}. Inspecting these phase diagrams, sketched in
Fig.~\ref{fig:exp1}e, it is obvious why droplets with larger initial
ethanol concentration have to loose more ethanol and are sustained
longer in \emph{stage1} before reaching the decomposition line that is
leading to the phase separation.

The small ethanol--rich droplets nucleating in the bulk of the mother
droplet coarsen quickly until a Janus state is reached. Due to the
swirl inside the mother droplet the coarsening droplets are
preferentially transported towards the rear side of the mother droplet
where they coalesce to even bigger droplets leading to one large
ethanol--rich droplet inside the mother droplet. When this coalesced
ethanol--rich droplet touches the inner surface of the mother droplet
it forms a three phase contact line and finally grows out of the
mother droplet forming a stable Janus-droplet when nucleation and
coarsening are completed, see Fig.~\ref{fig:exp1}c.


After the completed formation of self--propelling Janus droplets, they
consist of a water--rich leading droplet in contact with an
ethanol--rich trailing droplet that contains the majority of the
monoolein molecules, which were previously taken up by the mother
droplet. A water-rich droplet looses some of its volume (up to a
  few percent), while the Janus droplet velocity decreases
  exponentially until reaching a constant cruising speed.

No further nucleation in visible droplets or coarsening events could
be observed. However, estimating, the volume of both droplets and
comparing them to the droplet volume at the beginning of
\emph{stage~2} and the initial water/ethanol composition indicates
that a non--negligible amount of ethanol is still present in the
water--rich phase. Based on $\mu$PIV measurements of the flow field,
the stable Janus droplets can be identified as chains of neutral
swimmer~\cite{Suppmatt}, cf.~Fig.~\ref{fig:exp3}c. A surface flow at
the water--rich droplet can be directly observed in \emph{stage~3} by
secondary nucleated small ethanol--rich droplets at the surface of the
water--rich leading droplet. These small droplets are continuously
transported towards the ethanol--rich trailing droplet and merge with
it~\cite{Suppmatt}.
During transport the small droplets eventually grow in size,
reaching $1-5\,\rm\mu$m, when reaching the trailing ethanol
droplet~\cite{Suppmatt}.
Interestingly, the distribution of the secondary droplet surface
indicate the future macroscopic direction of motion of a Janus drop:
Each small droplet nucleated at the interface of the water--rich
droplet is a local sink for monoolein molecules in its near
surrounding and is thus influencing the symmetry of the Marangoni flow
pattern around the Janus droplet. If the distribution of the small
ethanol--rich droplets is homogeneous, the flow is symmetric with
respect to the symmetry axis of the Janus droplet leading to a
relatively straight trajectory.
In contrast, an inhomogeneous distribution of the secondary nucleated
droplets induces an asymmetric flow field and triggers a rotation of
the water--rich droplet away from the small droplets~\cite{Suppmatt}.

\subsubsection{Propulsion mechanism in stage 2 and 3}
\label{subsubsec:model}

Based on the above findings, we already concluded that the droplets in
\emph{stage~1} are driven by the solubilization of ethanol into the
surrounding oil phase. In view of the accumulation of monoolein inside
the trailing droplet in \emph{stage~3} and the continuing formation of
a monoolein depleted layer in the squalane around the leading droplet,
it suggests itself to model the propulsion in \emph{stage~2} and
\emph{3} from a balance between the work dissipated in viscous flows
and the chemical energy released by monoolein molecules on their path
from the ambient squalane phase into the trailing droplet. This
release of chemical energy drives a Marangoni flow that continuously
propels the Janus droplet. Since the small solvent molecules ethanol
and water diffuse much faster than the large monoolein molecules, we
assume a chemical equilibrium of water and ethanol between the two
bulk phases. Under these assumptions the concentration of solvents to
the bulk phases follows the binodal line of the phase diagram, while
the relative partition of ethanol and water to the leading and
trailing droplet is solely controlled by the total amount of monoolein
in the Janus droplets.

The flux of monoolein molecules into the trailing droplet in stage
\emph{3} can be expressed by the droplet velocity and the surface
coverage of monoolein on the interface of the leading droplet. Close
the final chemical equilibrium of the Janus droplet with the squalane
phase, it is justified to approximate the surface coverage $\Gamma$ by
the its corresponding equilibrium value. By virtue of the Gibbs
adsorption isotherm, we can relate the gradient of the interfacial
free energy to the gradient of the chemical potential of monoolein
molecules. The thin depletion layer of monoolein in the squalane
solution around the Janus droplet with thickness $\delta \ll R$
indicates large Peclet numbers $Pe=UR/D$, and we can set the chemical
potential of the adsorbed monoolein molecules at the tip of the
leading droplet to the value $\mu_\infty$ in the squalane
solution. The chemical potential at the three phase contact line can
be related to the concentration $n$ of monoolein molecules in the
trailing droplet. But as more and more monoolein accumulate in the
trailing droplet, the difference of chemical potentials of the
monoolein molecules in the squalane phase, $\mu_\infty$ and in the
ethanol--rich trailing droplet, $\mu$, decreases, and the Janus
droplets slow down. In return, this deceleration reduces the flux of
monoolein molecules into the trailing droplet and we predict an
exponential decrease of the droplet velocity $U(t)$ to zero in stage
\emph{3}. The time constant of this exponential decrease can be
expressed as
\begin{equation}
  \tau^{-1} = \frac{2\pi\,R\,\Gamma^2}{3\,C\,\eta\,V}
  \left.\frac{\partial\mu}{\partial n}\right|_{n=n_\infty}
  \label{eq:time_constant}
\end{equation}
where $\eta$ is the dynamic viscosity of the squalane phase, $V$ the
volume of the trailing droplet and $C$ a factor of order unity that
contains the dependence on the geometry of the Janus droplet and the
viscosity ratios of the bulk fluids \cite{Suppmatt}.

Spontaneous de--mixing of the mother droplet into an ethanol--rich and
a water--rich phase marking the end of \emph{stage 1} starts once the
composition of the Janus droplet reaches the boundary of the
miscibility gap. Provided a low initial ethanol concentration of the
mother droplet at the beginning of \emph{stage 1} we observe the
nucleation of small ethanol--rich minority phase droplets that quickly
coarsen. With a certain delay after the onset of nucleation, the first
minority phase droplets attach to the interface between the squalane
and the water--rich phase of the mother droplet. The thus formed
three--phase contact line allows the monoolein molecules on the
interface between the water--rich majority phase and the squalane
phase start to solubilize into the bulk of the ethanol--rich trailing
droplet. As outlined above for the droplet propulsion in \emph{stage
  3} the resulting gradient in the chemical potential of monoolein
molecules causes a Marangoni flow. Since the trailing droplet is
initially very small, it is quickly saturated with monoolein unless
larger, `fresh' minority phase droplets coalesce with the trailing
droplet and dilute the monoolein. The increase of concentration caused
by the influx of monoolein molecules from the interface of the leading
droplet is quickly surpassed by the dilution with minority phase
droplets that coalesce with the trailing droplet. As a result, the
difference of the chemical potential between the monoolein molecules
in the ambient squalane and the ethanol--rich trailing droplet
increases in \emph{stage 2}, and thus the propulsion velocity.

To estimate the balance of monooleine accumulation and its dilution in
\emph{stage 2}, we employ the average radius of the minority phase
droplets measured in our experiments. A quantitative analysis of the
experimental data shows that both the radius of the minority phase
droplets in the bulk of the mother droplet and the radius of the
trailing droplet in \emph{stage 2} conforms to a power law $\propto
t^{1.1}$ in time $t$ \cite{Suppmatt}. A linear growth of the typical
domain size after a quench into the miscibility gap of a binary liquid
was already predicted from dimensional analysis of the transport
equations \cite{Siggia1978} and observed in experiments
\cite{Mauri2003}.

Setting the growth exponent of the radius to $1$, the
non--dimensionalized undersaturation of monoolein in the trailing
droplet in \emph{stage 2}, and the propuslion velocity $U(t)\propto
x(t)$ are described by the equation
\begin{equation}
  \dot x =
  \frac{3}{t}\,\left(1-x\right)-\tau^{-1}\left(\frac{t_c}{t}\right)^3\,x~.
  \label{eq:early_time}
\end{equation}
Solutions $x(t)$ of Eqn.~(\ref{eq:early_time}) and, hence, the
velocity $U(t)$, displays a nearly linear rise in \emph{stage 2}, in
agreement with experimental findings \cite{Suppmatt}.

Since the final volume of the trailing droplet scales as $V \propto
R^3$, eqn.~(\ref{eq:time_constant}) predicts a relaxation time $\tau
\propto R^2$ in \emph{stage 3} whereas our experimental data show a
relaxation time that is independent on the dimensions of the droplet
\cite{Suppmatt}. In agreement with the experiments, the peak velocity
of self--propulsion is observed at a time $t=t_c$ marking the
transition between \emph{stage 2} and \emph{stage 3}. Numerical
solutions of eqn.~(\ref{eq:early_time}) show that the peak velocity is
monotoneously decreasing with an increasing ratio $t_c/\tau$ of the
coarsening time to the final relaxation time. Only in the limit of
small ratios $t_c/\tau \ll 1$ we find a constant peak velocity as in
the experiments \cite{Suppmatt}.

The model introduced above is based on the assumption that the water
and ethanol molecules of two fluid phases of the Janus droplet are
chemically equilibrated at any instance in time. However, we know from
experimental observations that this assumption cannot be completely
fulfilled. In \emph{stage 3}, in particular, we observe the nucleation
and growth of small ethanol--rich droplets at the interface between
the water--rich droplet with the squalane phase. These droplets are
advected to and merge with the trailing droplet. A secondary
nucleation indicates that the de--mixing of ethanol and water is still
ongoing in \emph{stage 3}. Droplets of the minority phase created in
this secondary nucleation could be responsible for an increased influx
of monoolein molecules into the trailing droplet that allows larger
droplets to reach a chemical equilibrium on the same time scale as
droplets with a smaller radius.

The discrepancy between the perdictions of the model and the
experimental results could also arise from an unjustified assumption
of chemcial equilibration of water and ethanol in the Janus
droplet. It is well possible that ethanol and water molecules in two
bulk phases of the Janus droplet are not close to a chemcial
equilibrium, and the gradients in the composition of the solvents also
contribute to the propelling Marangoni stresses. Moreover, the quick
uptake of monoolein in \emph{stage 1}, and the synchroneous appearance
of monodisperse minority phase droplets suggests a spinodal
decomposition of an unstable phase at the beginning of \emph{stage 2}
rather than a homogeneous nucleation. In view of the complexity of the
many competing kinetic effects, it does not come as a surprise that
the maximum velocity and the scaling of the exponential decay time
recorded in experiments do not conform to the prediction of our simple
simple model.

\subsubsection{Smart Carrier for Programmable DNA Cargo Delivery}
\label{subsubsec:smart_carrier}

\begin{figure}[c]
  \centering
  \includegraphics[width=15cm]{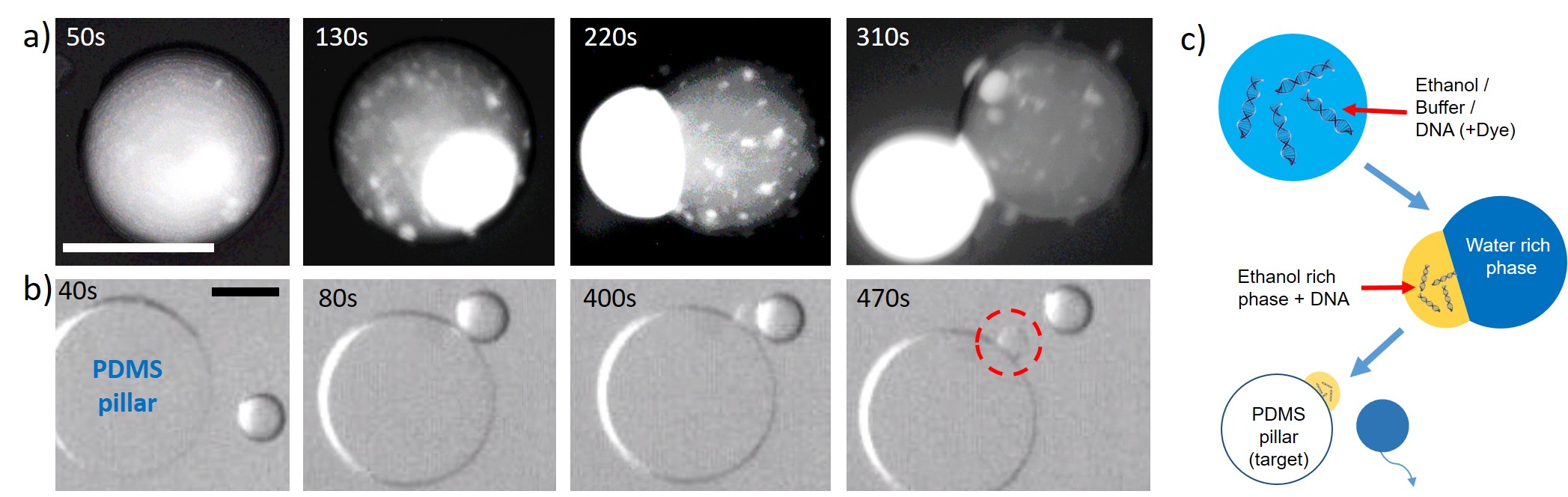}
  \caption{\textbf{DNA precipitation and cargo controlled delivery} a)
    Fluorescent microscopic time series showing the precipitation of
    fluorescent DNA into the ethanol--rich droplet, which finally
    pinches off.  b) Microscopic series of a DNA laden droplet moving
    near a PDMS obstacle. When touching the obstacle the droplet
    remains at the wall of the obstacle. The water--rich droplet is
    released from the pillar in \emph{stage~3} when the ethanol--rich
    droplet carrying the DNA is pinched off and spreads on the pillar
    surface.  c) Schematic of the process explained in a) and
    b). Note that precipitation of DNA in the ethanol-rich
      droplet only occurs in large presence of 0.3M sodium acetate.}
\label{fig:exp4}
\end{figure}

In this section, we will use the particular properties of our
evolutive self--propelling droplets for controlled cargo delivery in a
programmable manner. For that purpose, we used a water/ethanol
solution containing DNA ($\approx 0.1$~mg/ml) and $0.3$ M of sodium
acetate as droplet phase in a surrounding monoolein/squalane
solution. Even in the presence of DNA, the droplets undergo the
previously described evolution stages ending with a Janus droplet
consisting of a water--rich leading droplet and an ethanol--rich
trailing droplet. In contrast to similar bulk
methods~\cite{Zeugin1985, Smalla1993,Chiu2015}, DNA extraction in
self-propelling and phase separating droplets occurs in a single step
during less than $15$~minutes with sample volumes on the picoliter
scale, cf.~Fig.~\ref{fig:exp4}a,c.

To achieve a controlled cargo delivery, we use the fact that the
ethanol--rich droplet, which is formed during the evolutive
self--propulsion spreads easily on a polydimethylsyloxane (PDMS)
surface upon contact. Using this strategy, three different scenarios
of cargo delivery can be achieved simply by varying the initial
ethanol/water concentration of the droplet.

For intermediate ethanol concentration between $\approx 40~vol\%$ and
$\approx 60~vol\%$ the droplet is self--propelling $1--3$ minutes
before starting its phase separation.  After the Janus fromtion, a
\textit{drop-off} cargo delivery is realized; i.e.~the swimmer
delivers its cargo and leaves the target area.  If such a droplet
self--propels around an obstacle in the Janus \emph{stage~2} or $3$,
the ethanol--rich trailing droplet is spreading on the surface when
touching the obstacle, see Fig.~\ref{fig:exp4}b. This touching is
inevitable when the Janus droplet tries to swim away from the
obstacle. The remaining water--rich droplet is continuing its
self--propelled motion moving away from the obstacle. However, the
cargo can also be delivered if the droplet approaches the PDMS
obstacle already in \emph{stage 1}, or \emph{stage 2}, i.e. in a stage
before the trailing droplet is formed. In these stages a
self-propelling droplet will either stay in contact to the PDMS
surface, or self--propel in a bound orbit around it depending on its
specific squirmer mode~\cite{Papavassiliou2014}. Therefore it is just
a matter of time until the Janus droplet is formed leading to the
above described \textit{drop-off} cargo delivery.

For low ethanol concentration ($\approx~30~vol\%$), the formation of
Janus droplets occurs almost immediately after droplet production;
such a droplet has a short total running time. As consequence, those
water--rich droplets will stop to move soon after cargo delivery and
stay close to the cargo delivery area shielding it after a certain
number of delivery events.

For large ethanol concentration ($>70~vol\%$), the droplet is
self-propelling 10--15 minutes before starting it phase
separation. Several small water--rich droplets are formed inside an
ethanol--rich droplet that are finally pushed out of the larger
ethanol--rich droplet. The ethanol--rich droplets spread on the
surface after separation, while the water--rich droplets typically
continue to locomote, cf.~Fig.~\ref{fig:exp1}d. These self--propelling
water--rich droplets are still containing a non-negligible quantity of
ethanol being able to deliver the cargo. However this type of
multi-droplet cargo delivery is hardly controllable.

All these variants can be combined with selective DNA extraction, as
the DNA extraction is only enabled in the presence of $0.3$ M sodium
acetate.  Because the presence of salt reduced the DNA solubility in
water, and therefore the DNA precipitates and extracts into the
ethanol--rich trailing droplet during the phase separation.  It
results, that the final cargo content can be selectively controlled by
the initial chemical composition of the droplet. In particular, the
different value of delay time before the phase separation, are
controlling the brownian distance travel by the droplet before
starting the process of cargo-delivery. In total, three different
types of cargo delivery scenarios could be time controlled and
programmed just depending on the initial chemical
composition~\cite{Suppmatt}.

\subsection{Figures}

\section{Material and Method}
\label{sec:material_methods}

To fabricate the observation chambers, glass slides were cut into
squares of about ($2.5\times2.5$)~cm$^2$ and were pre-cleaned by
sonication in ethanol, acetone and toluene for 10~min each. The glass
squares were blow-dried with nitrogen gas after each cleaning
step. After this pre-cleaning they are immersed for 30~min in Piranha
etch (50~vol.\% sulfuric acid (98~\%) and 50~vol.\% hydrogen peroxide
(30~vol.\%). After that, the glass squares were thoroughly rinsed with
hot (80--90)$^{\circ}$ C ultrapure water and immersed in hot water for
another 15~min. Subsequently, the glass squares were again blow-dried
with nitrogen. The thus cleaned glass squares were coated with
octadecyl-trichlorosilane (OTS) by immersing them for 12~min in a
solution consisting of 50~ml bicyclohexane, 20--40 drops of carbon
tetrachloride, and 5--20 drops of OTS
\cite{Sagiv1980,Lessel2015}. Remains from the OTS solution were
removed from the OTS-coated glass by rinsing it with chloroform and
blow-drying with nitrogen. The OTS coated glass squares were stacked
with a cover slide (thickness 150~$\mu$m) as a spacer and glued
together (epoxy glue).

To conduct an experiment the observation chamber is prefilled with an
oil/surfactant mixture and ethanol/water mixture were carefully
injected into the reservoir using a glass capillary (20~$\mu$m inner
diameter, connected to a syringe pump. As water/ethanol were leaving
the glass capillary, droplets were formed spontaneously. The diameter
of the droplets could be controlled between 100 and 150~$\mu$m. The
density of the aqueous phase was always higher than that of the oily
phase (0.81~g/ml), even for the highest ethanol concentration
(80~vol.\%) used. The concentration of monoolein is fixed at 3.5~mg/ml
(i.e. 15~mM) if not explicitly mentioned otherwise. In experiments
with varying monoolein concentration, the ethanol/water concentration
was fixed to 50~vol.\%. All chemicals were purchased from
Sigma-Aldrich, except the fluorescent lipid that was purchase from
Avanti Polar Lipids.

\section{Conclusion}

\label{sec:conclusion}
We reported a general method to produce artificial self--propelling
Janus droplets, which evolve from monoolein stabilized water/ethanol
emulsion droplets. These type of active matter is universal and could
be achieved also for other surfactants and organic solvents including
liqueurs and brandies.  During ethanol solubilization, in
\emph{stage~1}, the droplet velocity fluctuates around a constant
value revealing a flow field of a weak pusher. The droplet takes up
monoolein molecules from the surrounding continuous phase finally
leading to a phase separation in \textit{stage~2}, where the droplet
velocity increases and the droplet moves as a neutral squirmer. When
the phase separation is completed, a Janus droplet is formed in
\emph{stage~3} with a leading water--rich droplet and a trailing
ethanol--rich droplet. The flow field around this Janus droplet is
that of a pair of neutral squirmers and the droplet velocity decays
exponentially with time. The chemical potential difference of the
monoolein molecules in the bulk of the oil phase and the ethanol phase
drives the motion in \emph{stage~2} and \emph{stage~3}. We introduced
a theoretical model that captures the relevant driving mechanism and
describes qualitatively the characteristic velocity behavior during
the \emph{stage~2} and \emph{stage~3}, despite the complexity of the
propulsion mechanism.

Due to the phase separation occurring in \emph{stage~2}, these
droplets could be used as a quasi-programmable smart
carrier. Macromolecules could be selectively precipitated and
delivered at a target location in a single step within a short time,
as was demonstrated for DNA delivery at PDMS pillars. Because the
droplets self--propelling properties are only controlled by the
initial chemical composition, we can realized three different cargo
delivery modes in a programmable manner. We expect that the simple
strategy reported here allows to design new classes of active
emulsions that could be used as smart self--propelling carriers or
operators. The combination of this controlled DNA cargo released with
microfluidic devices may enable new strategies for droplet digital
PCR~\cite{Hindson2013}, massive droplet DNA barcoding~\cite{Lan2016},
and gene, or drugs, delivery into cells~\cite{Chiu2015}.

\acknowledgments{ML, MB and RS acknowledges funding from SPP1726. The
  authors thank Frank Cichos, Klaus Kroy, Marcus Melke, Birthe
  Riechers, Dimitry Fedosov and Stephan Herminghaus for discussions.}


\end{document}